\documentclass[twocolumn,prl,amsmath,amssymb]{revtex4}
\usepackage{epsfig}
\usepackage{amsthm}
\usepackage{amssymb}
\usepackage{latexsym}
\usepackage{url}
\usepackage{amsmath, amscd}
\usepackage{verbatim}

\theoremstyle{definition}

\newcommand{\R}{\mathbb{R}}

\begin{document}

\title{On observation of position in quantum theory}
\author{A. \surname{Kryukov}} 
\affiliation{Department of Mathematics, University of Wisconsin Colleges, 34 Schroeder Ct, Madison, WI 53711, USA} 
\begin{abstract}
Newtonian and Scr{\"o}dinger dynamics can be formulated in a physically meaningful way within the same Hilbert space framework. This fact was recently used to discover an unexpected relation between classical and quantum motions that goes beyond the results provided by the Ehrenfest theorem. A formula relating the normal probability distribution and the Born rule was also found. Here the dynamical mechanism responsible for the latter formula is proposed and applied to measurements of macroscopic and microscopic systems. A relationship between the classical Brownian motion and the diffusion of state on the space of states is discovered. The role of measuring devices in quantum theory is investigated in the new framework. It is shown that the so-called collapse of the wave function is not measurement specific and does not require a ``concentration" near the eigenstates of the measured observable.  Instead, it is explained by the common diffusion of a state over the space of states under interaction with the apparatus and the environment. 
This in turn provides us with a basic reason for the definite position of macroscopic bodies in space.
\end{abstract}


\maketitle

\section{Introduction}

In a recent paper \cite{KryukovMacro} that serves as a foundation for the analysis presented here, an important new connection between Newtonian and Schr{\"o}dinger dynamics was derived. The starting point was a realization of classical and quantum mechanics within the same Hilbert space framework and identification of observables with vector fields on the sphere of normalized states. This resulted in a physically meaningful interpretation of components of the velocity of state that surpassed the Ehrenfest results on the motion of averages. Newtonian dynamics was shown to be the Schr{\"o}dinger dynamics of a system whose state is constrained to the classical phase space submanifold in the Hilbert space of states. This also resulted in a formula relating the normal probability distribution and the Born rule.

In this paper we continue to explore the implications of the proposed framework.
First, we show that there is a unique extension of Newtonian dynamics on the classical phase space submanifold to a unitary theory on the entire space of states. This allows us to find a connection between the Brownian motion of a macroscopic particle, the diffusion of state on the projective space $CP^{L_{2}}$ and the Born rule. It also allows us to make progress in understanding the process of measurement in quantum theory, the meaning of collapse of the wave function, the cause of the classicality of macroscopic bodies and the clarification of the role of decoherence in this.
To make the paper somewhat self-contained, we begin with a brief review of the results reported in \cite{KryukovMacro}.

\section{Newtonian and Schr{\"o}dinger dynamics in Hilbert space}

Macroscopic bodies have a well-defined position in space at any time. In quantum mechanics the state of a spinless particle with a known position ${\bf a} \in \R^{3}$ is described by the Dirac delta function $\delta^{3}_{\bf a}({\bf x})=\delta^{3}({\bf x}-{\bf a})$. The map $\omega: {\bf a} \longrightarrow \delta^{3}_{\bf a}$ provides a one-to-one correspondence between points ${\bf a} \in \R^{3}$ and state ``functions" $\delta^{3}_{\bf a}$. The set $\R^{3}$ can be then identified with the set $M_{3}$ of all delta functions in the space of state functions of the particle. 

The common Hilbert space $L_{2}({\R}^{3})$ of state functions of a particle does not contain delta functions.
By writing the inner product of functions $\varphi, \psi \in L_{2}(\R^{3})$ as
\begin{equation}
\label{innerdd}
(\varphi, \psi)_{L_{2}}=\int \delta^{3}({\bf x}-{\bf y})\varphi({\bf x}){\overline \psi}({\bf y})d^{3}{\bf x}d^{3}{\bf y}
\end{equation}
and approximating the kernel $\delta^{3}({\bf x}-{\bf y})$  with a Gaussian function, one obtains a new inner product in  $L_{2}(\R^3)$
\begin{equation}
\label{hh}
(\varphi, \psi)_{\bf H}=\int e^{-\frac{({\bf x}-{\bf y})^{2}}{8\sigma^{2}}}\varphi({\bf x})\overline{\psi}({\bf y})d^{3}{\bf x}d^{3}{\bf y}.
\end{equation}
The Hilbert space ${\bf H}$ obtained by completing $L_{2}(\R^3)$ with respect to this inner product contains delta functions and their derivatives.
In particular,
\begin{equation}
\label{delta1}
\int e^{-\frac{({\bf x}-{\bf y})^{2}}{8\sigma^{2}}}\delta^{3}({\bf x}-{\bf a})\delta^{3}({\bf y}-{\bf a})d^{3}{\bf x}d^{3}{\bf y}=1.
\end{equation}
It follows that the set $M_{3}$ of all delta functions  $\delta^{3}_{\bf a}({\bf x})$ with ${\bf a} \in \R^{3}$ form a submanifold of the unit sphere in the Hilbert space ${\bf H}$, diffeomorphic to $\R^3$. 

The map $\rho_{\sigma}: {\bf H} \longrightarrow L_{2}(\R^{3})$ that relates $L_{2}$ and ${\bf H}$-representations is given by the Gaussian kernel
\begin{equation}
\label{sigma}
\rho_{\sigma}({\bf x},{\bf y})=\left(\frac{1}{2\pi \sigma^{2}}\right)^{3/4}e^{-\frac{({\bf x}-{\bf y})^{2}}{4\sigma^{2}}}.
\end{equation}
In fact, multiplying the operators one can see that
\begin{equation}
\label{GG}
k({\bf x}, {\bf y})=(\rho^{\ast}_{\sigma}\rho_{\sigma})({\bf x}, {\bf y})=e^{-\frac{({\bf x}-{\bf y})^{2}}{8\sigma^{2}}},
\end{equation}
which is consistent with (\ref{hh}). 
The map $\rho_{\sigma}$ transforms delta functions $\delta^{3}_{\bf a}$ to Gaussian functions ${\widetilde \delta^{3}_{\bf a}}=\rho_{\sigma}(\delta^{3}_{\bf a})$, centered at ${\bf a}$. The image $M^{\sigma}_{3}$ of $M_{3}$ under $\rho_{\sigma}$ is an embedded submanifold of the unit sphere in $L_{2}(\R^{3})$ made of the functions ${\widetilde \delta^{3}_{\bf a}}$. The map $\omega_{\sigma}=\rho_{\sigma} \circ \omega: \R^3 \longrightarrow M^{\sigma}_{3}$ is a diffeomorphism. Here $\omega$ is the same as before. In what follows, the obtained realizations will be used interchangeably.

Let ${\bf r}={\bf a}(t)$ be a path with values in $\R^{3}$ and let $\varphi=\delta^{3}_{{\bf a}(t)}$ be the corresponding path in $M_{3}$. It is easy to see that the norm $\left \| \frac{d \varphi}{dt} \right \|^{2}_{H}$ of the velocity in the space ${\bf H}$ is given by
\begin{equation}
\label{parts1}
\left \| \frac{d \varphi}{dt} \right \|^{2}_{H}=
\left.\frac {\partial^{2}k({\bf x},{\bf y})}{\partial x^{i} \partial y^{k}}\right|_{{\bf x}={\bf y}={\bf a}} \frac {d{\bf a}^{i}}{dt}\frac {d{\bf a}^{k}}{dt}.
\end{equation}
Here  $k({\bf x},{\bf y})=e^{-\frac{({\bf x}-{\bf y})^{2}}{8\sigma^{2}}}$ as in (\ref{GG}), so that
\begin{equation}
\left.\frac {\partial^{2}k({\bf x},{\bf y})}{\partial x^{i} \partial y^{k}}\right|_{{\bf x}={\bf y}={\bf a}}=\frac{1}{4\sigma^{2}}\delta_{ik},
\end{equation}
where $\delta_{ik}$ is the Kronecker delta symbol. Assuming now that the distance in $\R^{3}$ is measured in the units of $2\sigma$, we obtain 
\begin{equation}
\label{Norms}
\left \| \frac{d \varphi}{dt} \right \|_{H}=\left \| \frac{d {\bf a}}{dt} \right \|_{\R^{3}}.
\end{equation}
It follows that the map $\omega: \R^{3} \longrightarrow {\bf H}$ is an isometric embedding. Furthermore, the set $M_{3}$ is complete in ${\bf H}$ so that there is no vector in ${\bf H}$ orthogonal to all of $M_{3}$. By defining the operations of addition $\oplus$ and multiplication by a scalar $\lambda \odot$  via $\omega({\bf a})\oplus\omega({\bf b})=\omega({\bf a}+{\bf b})$ and $\lambda \odot\omega({\bf a})=\omega(\lambda {\bf a})$ with $\omega$ as before, we obtain $M_{3}$ as a vector space isomorphic to the Euclidean space $\R^{3}$.

 The projection of velocity and acceleration of the state $\delta^{3}_{{\bf a}(t)}$ onto the Euclidean space $M_{3}$ yields correct Newtonian velocity and acceleration of the classical particle:
\begin{equation}
\label{v1}
\left( \frac{d}{dt}\delta^{3}_{{\bf a}}({\bf x}), -\frac{\partial}{\partial x^{i}} \delta^{3}_{{\bf a}}({\bf x})\right)_{\bf H} =\frac{d a^{i}}{dt}
\end{equation}
and 
\begin{equation}
\label{a11}
\left( \frac{d^{2}}{dt^{2}}\delta^{3}_{{\bf a}}({\bf x}) , -\frac{\partial}{\partial x^{i}} \delta^{3}_{{\bf a}}({\bf x})\right)_{\bf H} =\frac{d^{2} a^{i}}{dt^{2}}.
\end{equation}
The Newtonian dynamics of the classical particle now can be derived from the principle of least action for the action functional $S$ on paths in ${\bf H}$, defined by
\begin{equation}
\label{action}
\int k({\bf x},{\bf y})\left[\frac{m}{2}\frac{d \varphi_{t}({\bf x})}{dt} \frac{d{\overline  \varphi_{t}}({\bf y})}{dt}-V({\bf x}) \varphi_{t}({\bf x}) {\overline \varphi_{t}}({\bf y})\right]d^{3}{\bf x}d^{3}{\bf y}dt,
\end{equation}
where $m$ is the mass of the particle, $V$ is the potential and $k({\bf x}, {\bf y})=e^{-\frac{1}{2}({\bf x}-{\bf y})^{2}}$ (same as in (\ref{GG}) with $2\sigma=1$; see (\ref{Norms})). Namely, under the constraint $\varphi_{t}({\bf x})=\delta^{3}({\bf x}-{\bf a}(t))$ the action (\ref{action}) becomes
\begin{equation}
S=\int\left[\frac{m}{2}\left(\frac{d{\bf a}}{dt}\right)^{2}-V({\bf a})\right]dt,
\end{equation}
which is the classical action functional for the particle. 
This shows that a classical particle can be considered a constrained dynamical system with the state $\varphi$ of the particle and the velocity $\frac{d\varphi}{dt}$ of the state as dynamical variables. As shown in \cite{KryukovMacro}, a similar realization exists for mechanical systems consisting of any number of classical particles. 

Now that Newtonian dynamics is embedded in the framework of Hilbert spaces,  let's work from the opposite end and develop a vector representation in quantum theory. This representation will allow us to consider Newtonian and Schr{\"o}dinger dynamics on an equal footing.
The starting point is an identification of 
quantum observables with vector fields on the space of states. Given a self-adjoint operator ${\widehat A}$ on a Hilbert space $L_{2}$ of square-integrable functions (it could in particular be the tensor product space of a many body problem)  one can introduce the associated linear vector field $A_{\varphi}$ on $L_{2}$ by
\begin{equation}
\label{vector}
A_{\varphi}=-i{\widehat A}\varphi.
\end{equation}
If $D$ is the domain of the operator ${\widehat A}$, then $A_{\varphi}$ maps $D$ into the vector space $L_{2}$.
The commutator of observables and the commutator (Lie bracket) of the corresponding vector fields are related in a simple way:
\begin{equation}
\label{comm}
[A_{\varphi},B_{\varphi}]=[{\widehat A},{\widehat B}]\varphi.
\end{equation}
The field $A_{\varphi}$ associated with an observable, being restricted to the sphere $S^{L_{2}}$ of unit normalized states, is tangent to the sphere.

Under the embedding, the inner product on the Hilbert space $L_{2}$ yields the induced Riemannian metric on the sphere $S^{L_{2}}$. The projection $\pi: S^{L_{2}} \longrightarrow CP^{L_{2}}$ yields the induced Riemannian (Fubini-Study) metric on $CP^{L_{2}}$.
The resulting metrics can be used to find physically meaningful components of vector fields  $A_{\varphi}$ associated with observables. 
Since $A_{\varphi}$ is tangent to $S^{L_{2}}$, it can be decomposed into components tangent and orthogonal to the fibre $\{\varphi\}$ of the fibre bundle $\pi: S^{L_{2}} \longrightarrow CP^{L_{2}}$. These components have a simple physical meaning, justifying the use of the projective space $CP^{L_{2}}$ at this stage.
From
\begin{equation}
{\overline A} \equiv (\varphi, {\widehat A}\varphi)=(-i\varphi, -i{\widehat A}\varphi),
\end{equation}
one can see that the expected value of an observable ${\widehat A}$ in state $\varphi$ is the projection of the vector $-i{\widehat A}\varphi \in T_{\varphi}S^{L_{2}}$ onto the fibre $\{\varphi\}$.
Because
\begin{equation}
(\varphi, {\widehat A}^{2}\varphi)=({\widehat A}\varphi, {\widehat A}\varphi)=(-i{\widehat A}\varphi, -i{\widehat A}\varphi),
\end{equation}
the term $(\varphi, {\widehat A}^{2}\varphi)$ is the norm of the vector $-i{\widehat A}\varphi$ squared. The vector $-i{\widehat A}_{\bot}\varphi=-i{\widehat A}\varphi-(-i{\overline A}\varphi)$ associated with the operator ${\widehat A}-{\overline A}I$ is  orthogonal to the fibre $\{\varphi\}$.
Accordingly, the variance 
\begin{equation}
\Delta A^{2}=(\varphi, ({\widehat A}-{\overline A}I)^{2}\varphi)=(\varphi, {\widehat A}_{\bot}^{2}\varphi)=(-i{\widehat A}_{\bot}\varphi, -i{\widehat A}_{\bot}\varphi) 
\end{equation}
is the norm squared of the component $-i{\widehat A}_{\bot}\varphi$. 

From the Schr{\"o}dinger equation using the decomposition of $-i{\widehat h}\varphi$ onto the components parallel   and orthogonal to the fibre, we get
\begin{equation}
\label{evolll}
\frac{d\varphi}{dt}=-i{\overline E}\varphi+\left(-i{\widehat h}\varphi+i{\overline E}\varphi\right)=
-i{\overline E}\varphi-i {\widehat h}_{\perp}\varphi,
\end{equation}
where ${\overline E}$ is the expected value of the Hamiltonian ${\widehat h}$ in the state $\varphi$.
By projecting both sides of this equation by the differential $d\pi$ of the projection map $\pi: S^{L_{2}} \longrightarrow CP^{L_{2}}$, we obtain
\begin{equation}
\label{ddt}
\frac{d\{\varphi\}}{dt}=-i{\widehat h}_{\bot}\varphi.
\end{equation}
From this and the already derived equality
$
\|-i{\widehat h}_{\bot}\varphi\|=\Delta h,
$
it follows that the speed of evolution of state in the projective space is equal to the uncertainty of energy. 
This gives us two physically meaningful components of the velocity vector $\frac{d\varphi}{dt}$, corresponding to the expected value and uncertainty of the Hamiltonian. 

It turns out that the orthogonal component $-i{\widehat h}_{\bot}\varphi$ of the velocity can also be decomposed into physically meaningful quantities.
More importantly, the embedding $\omega_{\sigma}=\rho_{\sigma} \circ \omega$ of $\R^{3}$ into the space of states $L_{2}(\R^3)$ together with the vector representation of observables provide us  with a bridge between Newtonian to Schr{\"o}dinger dynamics. 
To demonstrate this, recall first that the basic relation between the classical and quantum physics is given by the Ehrenfest theorem
\begin{equation}
\label{Her}
\frac{d}{dt}(\varphi, {\widehat A}\varphi)=-i(\varphi, [{\widehat A}, {\widehat h}]\varphi).
\end{equation}
Here ${\widehat A}$ does not depend on $t$. 
Compare (\ref{Her})  to another equation that follows from the Schr{\"o}dinger dynamics:
\begin{equation}
\label{projj}
 2\left(\frac{d  \varphi}{dt}, -i {\widehat A} \varphi \right)=
 \left( \varphi, \{{\widehat A}, {\widehat h} \}\varphi \right)-\left(\varphi,[{\widehat A}, {\widehat h}]\varphi\right).
\end{equation}
The Ehrenfest theorem (\ref{Her}) for a time-independent observable amounts to using  the imaginary part of (\ref{projj}), i.e., the part with the commutator $[{\widehat A}, {\widehat h}]$. The left hand side of (\ref{projj}) is twice the projection of the velocity of state onto the vector field associated with the observable ${\widehat A}$. The real part of this projection (the term with the anticommutator $\{{\widehat A}, {\widehat h}\}$) is twice the projection in the sense of Riemannian metric on $S^{L_{2}}$. This Riemannian projection will be now used to identify components of the velocity of state.

Suppose that at $t=0$, a microscopic particle is prepared in the state
\begin{equation}
\label{initial}
    \varphi_{0}({\bf x})=\left(\frac{1}{2\pi\sigma^{2}}\right)^{3/4}e^{-\tfrac{({\bf x}-{\bf x}_{0})^{2}}{4\sigma^{2}}}e^{i\tfrac{{\bf p}_{0}({\bf x}-{\bf x}_{0})}{\hbar}},
\end{equation}
where $\sigma$ is the same as in (\ref{sigma}) and ${\bf p}_{0}=m{\bf v}_{0}$ with ${\bf v}_{0}$ being the initial group-velocity of the packet. 
The set $M^{\sigma}_{3,3}$ of all initial states $\varphi_{0}$ given by (\ref{initial}) form a $6$-dimensional embedded submanifold in $L_{2}(\R^{3})$. 
Consider the set of all fibres of the bundle $\pi: S^{L_{2}} \longrightarrow CP^{L_{2}}$ through the points of $M^{\sigma}_{3,3}$. The resulting bundle $\pi: M^{\sigma}_{3,3}\times S^{1} \longrightarrow M^{\sigma}_{3,3}$ identifies $M^{\sigma}_{3,3}$ with a submanifold of $CP^{L_{2}}$, denoted by the same symbol.
The map $\Omega: \R^{3}\times \R^{3} \longrightarrow M^{\sigma}_{3,3}$,
\begin{equation}
\label{omega}
\Omega({\bf a},{\bf p})=\left(\frac{1}{2\pi\sigma^{2}}\right)^{3/4}e^{-\frac{({\bf x}-{\bf a})^{2}}{4\sigma^{2}}}e^{i\frac{{\bf p}({\bf x}-{\bf a})}{\hbar}}
\end{equation} 
is a diffeomorphism from the classical phase space of the particle onto $M^{\sigma}_{3,3}$. 
For $\Omega=r e^{i\theta}$, where $r$ is the modulus and $\theta$ is the argument of $\Omega$, the vectors $\frac{\partial r}{\partial x^{\alpha}}e^{i\theta}$ and $i\frac{\partial \theta}{\partial p^{\beta}}re^{i\theta}$ are orthogonal to the fibre $\{\varphi_{0}\}$ at the point $\varphi_{0}$ in $L_{2}(\R^3)$. These vectors can be then identified with vectors tangent to $M^{\sigma}_{3,3}$, considered as a submanifold of $CP^{L_{2}}$. They form a basis in the tangent space $T_{\{\varphi_{0}\}}M^{\sigma}_{3,3}$. Furthermore, the induced Riemannian metric is the usual Fubini-Study metric on $CP^{L_{2}}$, restricted to $M^{\sigma}_{3,3}$. 

For any path $\{\varphi\}_{\tau}$ with values in $M^{\sigma}_{3,3}\subset CP^{L_{2}}$ the norm of velocity vector $\frac{d \{\varphi\}}{d\tau}$ in the Fubini-Study metric is given by
\begin{equation}
\label{phaseMetric}
\left\|\frac{d \{\varphi\}}{d \tau}\right\|^{2}_{FS}=\frac{1}{4\sigma^{2}}\left\|\frac{d {\bf a}}{d\tau}\right\|^{2}_{\R^{3}}+\frac{\sigma^{2}}{\hbar^{2}}\left\|\frac{d {\bf p}}{d\tau}\right\|^{2}_{\R^{3}}.
\end{equation}
It follows that under a proper choice of units, the map $\Omega$ is an isometry that identifies the Euclidean phase space $\R^{3}\times \R^{3}$ of the particle with the submanifold $M^{\sigma}_{3,3} \subset CP^{L_{2}}$ furnished with the induced metric. The map $\Omega$ is an extension to the phase space of the isometric embedding $\omega_{\sigma}=\rho_{\sigma}\circ\omega$ of the space $\R^{3}$, considered earlier in this section.


Let us now decompose the orthogonal component $-\frac{i}{\hbar}{\widehat h}_{\perp}\varphi$ of the velocity $\frac{d\varphi}{dt}$. 
Calculation of the projection of the velocity $\frac{d \varphi}{dt}$ onto the unit vector $-\widehat{\frac{\partial r}{\partial x^{\alpha}}}e^{i\theta}$ (i.e., the classical space component of $\frac{d\varphi}{dt}$) for an arbitrary Hamiltonian of the form ${\widehat h}=-\frac{\hbar^{2}}{2m}\Delta+V({\bf x})$ yields
\begin{equation}
\label{pproj}
\left.\mathrm{Re}\left(\frac{d \varphi}{dt}, -\widehat{ \frac{\partial r}{\partial x^{\alpha}}}e^{i\theta}\right)\right|_{t=0}=\left.\left(\frac{d r}{dt}, -\widehat{ \frac{\partial r}{\partial x^{\alpha}}}\right)\right|_{t=0}=\frac{v^{\alpha}_{0}}{2\sigma}.
\end{equation}
Calculation of the projection of velocity $\frac{d \varphi}{dt}$ onto the unit vector $i\widehat{\frac{\partial\theta}{\partial p^{\alpha}}}\varphi$ (momentum space component) gives
\begin{equation}
\label{w}
\left.\mathrm{Re} \left(\frac{d\varphi}{dt}, i\widehat{\frac{\partial\theta}{\partial p^{\alpha}}}\varphi\right)\right|_{t=0}=\frac{mw^{\alpha} \sigma}{\hbar},
\end{equation}
where
\begin{equation}
mw^{\alpha}=-\left.\frac{\partial V({\bf x})}{\partial x^{\alpha}}\right|_{{\bf x}={\bf x}_{0}}
\end{equation}
and $\sigma$ is assumed to be small enough for the linear approximation of $V({\bf x})$ to be valid within intervals of length $\sigma$. 

The velocity $\frac{d\varphi}{dt}$ also contains component due to the change in $\sigma$ (spreading), which is orthogonal to the fibre $\{\varphi\}$ and the phase space $M^{\sigma}_{3,3}$, and is equal to
\begin{equation}
\label{spreadcomp}
\left.\mathrm{Re} \left (\frac{d\varphi}{dt}, i\widehat{\frac{d\varphi}{d\sigma}}\right)\right|_{t=0}=\frac{\sqrt{2}\hbar}{8\sigma^{2}m}.
\end{equation}
Calculation of the norm of $\frac{d\varphi}{dt}=\frac{i}{\hbar}{\widehat h}\varphi$ at $t=0$ gives
\begin{equation}
\label{decomposition}
\left\|\frac{d\varphi}{dt}\right\|^{2}=\frac{{\overline E}^{2}}{\hbar^{2}}+\frac{{\bf v}^{2}_{0}}{4\sigma^{2}}+\frac{m^{2}{\bf w}^{2}{\sigma}^{2}}{\hbar^{2}}+\frac{\hbar^{2}}{32\sigma^{4}m^{2}},
\end{equation}
which is the sum of squares of the components. This completes a decomposition of the velocity of state at any point $\varphi_{0} \in M^{\sigma}_{3,3}$. 

From (\ref{pproj}) and (\ref{w}) and a simple consistency check, one can see that the phase space components of the velocity of state $\frac{d\varphi}{dt}$ assume correct classical values at any point $\varphi_{0}  \in M^{\sigma}_{3,3}$. This remains true for the time dependent potentials as well. The immediate consequence of this and the linear nature of the Schr{\"o}dinger equation is that under the Schr{\"o}dinger evolution with the Hamiltonian ${\widehat h}=-\frac{\hbar^{2}}{2m}\Delta+V({\bf x},t)$, the state constrained to $M^{\sigma}_{3,3}\subset CP^{L_{2}}$ moves like a point in the phase space representing a particle in Newtonian dynamics. 
More generally, 
Newtonian dynamics of $n$ particles is the Schr{\"o}dinger dynamics of $n$-particle quantum system whose state is constrained to the phase-space submanifold $M^{\sigma}_{3n,3n}$ of the projective space for $L_{2}(\R^{3})\otimes \ ... \ \otimes L_{2}(\R^{3})$, formed by tensor product states $\varphi_{1}\otimes \ ... \ \otimes \varphi_{n}$ with $\varphi_{k}$ of the form (\ref{initial}). 

Note that this result is precise, as long as the width $\sigma$ is sufficiently small, so that the linear approximation of the potential on intervals of length $\sigma$ is valid. This latter condition is not so much a restriction, as the choice of $\sigma$ is in our hands.  Starting with potentials with the highest second derivative available to us in practice we can always choose $\sigma$ small enough to satisfy the classical behavior. That is, all quantum effects in the system with the state constrained to the phase space submanifold with the appropriate value of $\sigma$ will disappear. 

Note again that the velocity and acceleration terms in (\ref{decomposition}) are orthogonal to the fibre $\{\varphi_{0}\}$ of the fibration $\pi: S^{L_{2}}\longrightarrow CP^{L_{2}}$, showing that these Newtonian variables have to do with the motion in the projective space $CP^{L_{2}}$. The velocity of spreading is also orthogonal to the fibre. Later in the paper this will be related to the fact that ``collapse" of the wave function is also happening in the projective space.

To complete a review of \cite{KryukovMacro}, note that isometric embedding of the classical space $M^{\sigma}_{3}$ into the space of states $L_{2}(\R^3)$ results in a relationship between distances in $\R^3$ and the projective space $CP^{L_{2}}$.
The distance between two points ${\bf a}$ and  ${\bf b}$ in $\R^{3}$ is $\left\|{\bf a}-{\bf b}\right\|_{\R^{3}}$. Under the embedding of the classical space into the space of states, the variable ${\bf a}$ is represented by the state $\tilde{\delta}^{3}_{\bf a}$. The set of states $\tilde{\delta}^{3}_{\bf a}$ form a submanifold $M^{\sigma}_{3}$ in the Hilbert spaces of states $L_{2}(\R^{3})$, which is "twisted" in $L_{2}(\R^{3})$. It belongs to the sphere $S^{L_{2}}$ and spans all dimensions of $L_{2}(\R^{3})$. The distance between the states $\tilde{\delta}^{3}_{\bf a}$, $\tilde{\delta}^{3}_{\bf b}$ on the sphere $S^{L_{2}}$ or in the projective space $CP^{L_{2}}$ is not equal to $\left\|{\bf a}-{\bf b}\right\|_{\R^{3}}$. In fact, the former distance measures length of a geodesic between the states while the latter is obtained using the same metric on the space of states, but applied along a geodesic in the twisted manifold $M^{\sigma}_{3}$. 
The precise relation between the two distances is given by
\begin{equation}
\label{main}
e^{-\frac{({\bf a}-{\bf b})^{2}}{4\sigma^{2}}}=\cos^{2}\rho(\tilde{\delta}^{3}_{\bf a}, \tilde{\delta}^{3}_{\bf b}),
\end{equation}
where $\rho$ is the Fubini-Study distance between states in $CP^{L_{2}}$. The projective space of states $CP^{L_{2}}$ appears here for a good reason: the fibres of the fibration $\pi: S^{L_{2}} \longrightarrow CP^{L_{2}}$ through the points of the classical space $M^{\sigma}_{3}$ are orthogonal to this space. This is why the distance in  $M^{\sigma}_{3}$ can be expressed in terms of the distance in $CP^{L_{2}}$.

The relation (\ref{main}) has an immediate implication onto the form of probability distributions of random variables over $M^{\sigma}_{3}$ and $CP^{L_{2}}$.
In particular, consider a random variable $\psi$ over $CP^{L_{2}}$. Suppose that the restricted random variable $\psi$ defined over $M^{\sigma}_{3}=\R^3$  is distributed normally on $\R^3$. (That is, the truncated distribution is normal.) Then, by (\ref{main}), the isotropic (i.e., direction independent at each point) probability distribution of $\psi$ over $CP^{L_{2}}$ must satisfy the Born rule. That is,
the normal probability distribution of a position random variable for a particle in the classical space implies the Born rule for transitions between arbitrary quantum states of the particle and vice versa.

\section{Extension of Newtonian dynamics to the space $CP^{L_{2}}$ of quantum states}

Recall that the Schr{\"o}dinger dynamics with the Hamiltonian ${\widehat h}=-\frac{\hbar^{2}}{2m}\Delta+V({\bf x})$ was used to find the classical and momentum space components of the velocity $\frac{d\varphi}{dt}$ for a particle. For convenience, these results (formulae (\ref{pproj}) and (\ref{w})) are reproduced here: 
\begin{equation}
\label{pproj1}
\left.\mathrm{Re}\left(\frac{d \varphi}{dt}, -\widehat{ \frac{\partial r}{\partial x^{\alpha}}}e^{i\theta}\right)\right|_{t=0}=\left.\left(\frac{d r}{dt}, -\widehat{ \frac{\partial r}{\partial x^{\alpha}}}\right)\right|_{t=0}=\frac{v^{\alpha}}{2\sigma},
\end{equation}
and
\begin{equation}
\label{w1}
\left.\mathrm{Re} \left(\frac{d\varphi}{dt}, i\widehat{\frac{\partial\theta}{\partial p^{\alpha}}}\varphi\right)\right|_{t=0}=\frac{mw^{\alpha} \sigma}{\hbar},
\end{equation}
where
\begin{equation}
\label{linearA}
mw^{\alpha}=-\left.\frac{\partial V({\bf x})}{\partial x^{\alpha}}\right|_{{\bf x}={\bf a}}
\end{equation}
and $\sigma$ is sufficiently small for the linear approximation of $V({\bf x})$ to be valid over intervals of length $\sigma$. Formulae (\ref{pproj1}) and (\ref{w1}) were used to establish that Newtonian dynamics of a particle is the Schr{\"o}dinger dynamics constrained to the classical phase space $M^{\sigma}_{3,3}$ of the particle. 

Suppose on the contrary that for any initial state  $\varphi_{0}$ of the form
\begin{equation}
\label{del1}
\varphi_{0}({\bf x})=
\left(\frac{1}{2\pi \sigma^{2}}\right)^{3/4}e^{-\frac{({\bf x}-{\bf a})^{2}}{4\sigma^{2}}}e^{i\frac{{\bf p}({\bf x}-{\bf a})}{\hbar}}
\end{equation}
there exists a path $\varphi=\varphi_{t}$  in $L_{2}(\R^{3})$, passing at $t=0$ through the point $\varphi_{0}$, and such that  (\ref{pproj1}) and (\ref{w1}) are satisfied.
Suppose further that the evolution $\varphi=\varphi_{t}$ is unitary and, therefore, by Stone's theorem, 
\begin{equation}
\label{back}
\frac{d\varphi}{dt}=-\frac{i}{\hbar} {\widehat H}\varphi
\end{equation}
for a Hermitian operator ${\widehat H}$ on $L_{2}(\R^{3})$.
The claim is that this implies the Ehrenfest theorem on states (\ref{del1}) and, by linear extension, on the entire space $L_{2}(\R^{3})$. It then follows that the operator ${\widehat H}$ is uniquely defined and is equal to $-\frac{\hbar^{2}}{2m}\Delta+V({\bf x})$. More precisely, 

{\it There is a unique unitary evolution on $L_{2}(\R^3)$ such that under this evolution the system constrained to the classical phase space $M^{\sigma}_{3,3}$ satisfies Newtonian equations of motion for the particle. This evolution obeys the Schr{\"o}dinger equation of motion with the usual Hamiltonian ${\widehat h}=-\frac{\hbar^{2}}{2m}\Delta+V({\bf x})$.}

Let's first prove that (\ref{pproj1}) and (\ref{w1}) imply the Ehrenfest theorem on states $\varphi \in M^{\sigma}_{3,3}$. As discussed, the Ehrenfest theorem can be written in the following form:
\begin{equation}
\label{EE1}
2\mathrm{Re} \left(\frac{d\varphi}{dt}, {\widehat x} \varphi \right)=\left(\varphi, \frac{\widehat p}{m}\varphi \right)
\end{equation}
and
\begin{equation}
\label{EE2}
2\mathrm{Re} \left(\frac{d\varphi}{dt}, {\widehat p} \varphi \right)=\left(\varphi, - \nabla V({\bf x}) \varphi \right).
\end{equation}
From (\ref{pproj1}) and (\ref{del1}) we have at $t=0$,
\begin{equation}
\label{pproj2}
\frac{v^{\alpha}}{2\sigma}=\mathrm{Re}\left(\frac{d \varphi}{dt}, -\widehat{ \frac{\partial r}{\partial x^{\alpha}}}e^{i\theta}\right)=\frac{1}{\sigma}\mathrm{Re}\left(\frac{d\varphi}{dt}, (x-a)^{\alpha}\varphi\right).
\end{equation}
Because of the unitary condition, we have $\mathrm{Re}\left(\frac{d\varphi}{dt}, \varphi\right)=0$ and so (\ref{pproj2}) yields
\begin{equation}
\label{pproj3}
2\mathrm{Re}\left(\frac{d\varphi}{dt}, x^{\alpha}\varphi\right)=v^{\alpha}=\frac{p^{\alpha}}{m}.
\end{equation}
Together with 
$
\left(\varphi, {\widehat p}\varphi\right)=\left(\varphi, {\bf p}\varphi\right)={\bf p}
$
this gives the first Ehrenfest theorem (\ref{EE1}) on states $\varphi \in M^{\sigma}_{3,3}$.

Similarly, from (\ref{w1}) and (\ref{del1}) we have at $t=0$,
\begin{equation}
\label{w2}
\frac{mw^{\alpha} \sigma}{\hbar}=\mathrm{Re} \left(\frac{d\varphi}{dt}, i\widehat{\frac{\partial\theta}{\partial p^{\alpha}}}\varphi\right)=\frac{\hbar}{\sigma}\mathrm{Re}\left(\frac{d\varphi}{dt}, \frac{i(x-a)^{\alpha}}{\hbar}\varphi\right),
\end{equation}
with
\begin{equation}
mw^{\alpha}=-\left.\frac{\partial V({\bf x})}{\partial x^{\alpha}}\right|_{{\bf x}={\bf a}}.
\end{equation}
On the other hand, 
\begin{equation}
{\widehat p}\varphi=-i\hbar \nabla \varphi=-i\hbar\left(-\frac{{\bf x}-{\bf a}}{2\sigma^{2}}+\frac{i{\bf p}}{\hbar}\right)\varphi.
\end{equation}
Again, from the unitary condition we have $\mathrm{Re}\left(\frac{d\varphi}{dt}, \varphi\right)=0$ and so we can rewrite (\ref{w2}) as 
\begin{equation}
\label{w3}
\frac{mw^{\alpha} \sigma}{\hbar}=\frac{\sigma}{\hbar}\mathrm{Re}\left(\frac{d\varphi}{dt}, {\widehat p}^{\alpha}\varphi \right),
\end{equation}
or,
\begin{equation}
\label{w4}
2\mathrm{Re}\left(\frac{d\varphi}{dt}, {\widehat p}^{\alpha}\varphi \right)=mw^{\alpha}.
\end{equation}
From this and (\ref{linearA}), we get the second Ehrenfest theorem (\ref{EE2}) on states $\varphi \in M^{\sigma}_{3,3}$. From (\ref{pproj2}) and (\ref{w2}), one can also see that velocity and acceleration terms are the real and imaginary parts of a complex vector, tangent to $M^{\sigma}_{3,3}$. 

Now, from the derived Ehrenfest theorems and the Stone's theorem for unitary evolution
\begin{equation}
\label{back1}
\frac{d\varphi}{dt}=-\frac{i}{\hbar} {\widehat H}\varphi,
\end{equation}
we get the following equations for the unknown Hermitian operator ${\widehat H}$, valid for all functions $\varphi$ in $M^{\sigma}_{3,3}$:
\begin{equation}
\label{a}
\left( \varphi, i[{\widehat H},{\widehat x}]\varphi \right)=\frac{\hbar }{m}\left(\varphi, {\widehat p}\varphi\right)
\end{equation}
and
\begin{equation}
\label{b}
\left( \varphi, i[{\widehat H},{\widehat p}]\varphi \right)=\hbar\left(\varphi, -\nabla V({\bf x})\varphi\right).
\end{equation} 
The operators on the right hand sides of (\ref{a}) and (\ref{b}) are so far defined only on $M^{\sigma}_{3,3}$. 
However, because the set $M^{\sigma}_{3,3}$ is complete in $L_{2}(\R^{3})$, there is a unique linear extension of the operators to (a dense subset of) $L_{2}(\R^{3})$. Therefore, there is a unique extension of the right hand sides of (\ref{a}) and (\ref{b}) to quadratic forms on $L_{2}(\R^{3})$.
Let us show that there is a unique operator ${\widehat H}$ for which the equations (\ref{a}) and (\ref{b}) remain true for these extensions. That is, there exists a unique operator ${\widehat H}$ for  which
\begin{equation}
\label{a1}
\left( f, i[{\widehat H},{\widehat x}]f \right)=\frac{\hbar }{m}\left(f, {\widehat p}f\right)
\end{equation}
and
\begin{equation}
\label{b1}
\left( f, i[{\widehat H},{\widehat p}]f \right)=\hbar\left(f, -\nabla V({\bf x})f\right)
\end{equation}
for all functions $f$ in the dense subset $D$ of $L_{2}(\R^{3})$, which is the common domain of all involved operators.
In fact, by choosing an orthonormal basis $\{e_{j}\}$ in $D$ and considering (\ref{a1}), (\ref{b1}) on functions $f=e_{k}+e_{l}$ and $f=e_{k}+ie_{l}$ we conclude that all matrix elements of the operators on the left and right of the equations (\ref{a1}) and  (\ref{b1}) must be equal. So the equations can be written in the operator form
\begin{equation}
\label{1}
i[{\widehat H},{\widehat x}]=\frac{\hbar }{m}{\widehat p}
\end{equation}
and
\begin{equation}
\label{2}
i[{\widehat H},{\widehat p}]=-\hbar\nabla V({\bf x}).
\end{equation}
From (\ref{1}) and (\ref{2}) it then follows that, up to an irrelevant constant, ${\widehat H}=\frac{{\widehat p}^{2}}{2m}+V({\bf x})$. 

The same result holds true for time-dependent potentials $V({\bf x}, t)$ as well. Generalization to the case of $n$ interacting distinguishable particles described by tensor product of states (\ref{del1}) is straightforward and leads to the Hamiltonian 
${\widehat H}=\sum_{k}\frac{{\widehat p_{k}}^{2}}{2m_{k}}+V({\bf x}_{1}, ..., {\bf x}_{n})$. (See \cite{KryukovMacro} for the analogues of (\ref{pproj1}) and (\ref{w1}) in this case. Otherwise the derivation for the single particle operators ${\widehat x}_{k}$, ${\widehat p}_{k}$ mimics the one given here.)

By (\ref{omega}), a point $\varphi_{0}$ in the classical phase space  $M^{\sigma}_{3,3}$ defines the initial position and velocity of the particle in $\R^3$. The solution of Newton's equations with this initial condition defines a unique classical path $({\bf a}_{t}, {\bf p}_{t})$ of the particle. Let's call the (non-linear) operator $U_{c}(t,0): M^{\sigma}_{3,3} \longrightarrow M^{\sigma}_{3,3}$, given by  
\begin{equation}
U_{c}(t,0)\left( \Omega({\bf a}_0, {\bf p}_0)\right)=\Omega({\bf a}_t, {\bf p}_t)
\end{equation}
with $\Omega$ given by (\ref{omega}), the {\it Newtonian evolution operator}. It was shown that there exists a unique unitary evolution operator  $U_{q}(t,0): L_{2}(\R^3) \longrightarrow L_{2}(\R^3)$, 
such that  $U_{q}(t,0)\varphi_{0}=\varphi_{t}$ satisfies (\ref{pproj1}), (\ref{w1}) for all $\varphi_{0} \in M^{\sigma}_{3,3}$. The operator $U_{q}(t,0)$ will be called the CQ-extension of the operator $U_{c}(t,0)$. It turned out to be the usual Schr{\"o}dinger evolution operator. The domain $L_{2}(\R^3)$ of this operator is the (closure of the) linear envelop of the domain $M^{\sigma}_{3,3}$ of the Newtonian evolution operator. The component of the velocity vector field $\frac{d U_{q}(t,0)\varphi_{0}}{dt}$ tangent to $M^{\sigma}_{3,3}$ gives the usual Newtonian velocity and acceleration of the particle. The meaning of the additional components of $\frac{d \varphi}{dt}$ was revealed in (\ref{decomposition}).

 The obtained embedding of the classical phase space  into the space of states complemented by existence and uniqueness of extension of Newtonian to Schr{\"o}dinger evolution signifies that Newtonian dynamics found its full-fledged realization within the realm of quantum physics governed by the Schr{\"o}dinger equation. This realization is valid independently of whether it is taken to mean the actual physical embedding or only as a mathematical representation. In either case the derived relationship of classical and quantum dynamics can be used to generate other valid and physically meaningful results, some of which are obtained below.

\section{Observation of position of macroscopic and microscopic particles}

The goal now is to understand the relationship of measurements in classical and quantum mechanics. This will be done by using measurement of position as a rather general example. A powerful tool available to us is the bridge between classical and quantum dynamics, as constructed in the previous sections. To repeat, we know that the Schr{\"o}dinger dynamics restricted to the classical phase space yields the Newtonian dynamics. Likewise, Newtonian dynamics of a particle extends in a unique way to the Schr{\"o}dinger dynamics on the space of states, via the CQ-extension. We also know that under the isotropy condition, the normal probability distribution on $\R^3$ extends uniquely to $CP^{L_{2}}$ and yields the Born rule.

To relate the measurements, we can use a model of measurement {\it consistent with Newtonian dynamics and the normal distribution of measurement results}. The dynamics of any such model has a unique CQ-extension to Schr{\"o}dinger dynamics of the corresponding quantum model. This unique extension is guaranteed to be consistent with the Born rule for transitions between quantum states. Because the extension was proved to be unique, the main assumption that is made here is that {\it the underlying physical process responsible for measurements on macroscopic and microscopic systems  is fundamentally the same.} Consequently, the (unique) CQ-extension of dynamics of an appropriate classical model of measurement must be adequate for dealing with measurements of a microscopic  system.

To our advantage, the position of a macroscopic particle under observation generically satisfies the normal distribution law. This is consistent with the central limit theorem and indicates that a specific way in which position is measured is not important. From this and the constructed extension of the normal distribution, it follows that the Born rule on the space of states must be as generic as the normal distribution in macro-physics, which is consistent with observation.

One common way of finding the position of a macroscopic particle is to expose it to light of sufficiently short wavelength and to observe the scattered photons. In many cases, due to the unknown path of the incident photons, multiple scattering events on the particle, random change in position of the particle, etc., the process of observation can be described by the diffusion equation with the observed position of the particle experiencing Brownian motion from an initial point during the time of observation. 
Within the applicability of geometric optics, such a measuring experiment is consistent with Newtonian dynamics. It is also consistent with the normal distribution of observed position of the particle.

Let us use this particular experiment and analyze properties of the CQ-extension of the underlying dynamics. (See \cite{Sanz} for possible ways of describing the classical Brownian motion using Hamiltonians.) 
Thus, we assume that position of the measured macroscopic particle in the experiment is a random variable on $\R^3$ and that Einstein's derivation of Brownian motion for the particle is valid. Under the CQ-extension of Brownian motion the state $\varphi$ of the particle  is a random variable on the sphere $S^{L_{2}}$.
In the non-relativistic quantum mechanics a particle cannot disappear or be created.  From unitarity of the extension we know that the state of the measured particle cannot disappear either. It simply evolves along the unit sphere in the space of states. To express this conservation of states mathematically, consider the density of states functional $\rho_{t} [\varphi;\psi]$. This functional measures the number of states of particles in the space of states on a neighborhood of a point $\varphi \in L_{2}(\R^3)$ at time $t$, given the initial state $\psi$ of the particle. 
Under the isometry $\omega: \R^3 \longrightarrow M^{\sigma}_{3} \subset L_{2}(\R^3)$ the states in $M^{\sigma}_{3}$ are identified with (positions of) particles. So the density of states functional $\rho_{t}[\varphi;\psi]$ must be an extension of the density of particles function $\rho_{t}({\bf a}; {\bf b})$ in the usual diffusion on $\R^{3}$. Here ${\bf b}$ is the initial position of the particle.  In other words, $\rho_{t}({\bf a}; {\bf b})=\rho_{t}[{\tilde \delta}^{3}_{\bf a}; {\tilde \delta}^{3}_{\bf b}]$. 
The precise meaning of the functional $\rho_{t}[\varphi;\psi]$ for the measurements resulting in a specific position of a microscopic particle will be explained below.
From the conservation of states, we have:
\begin{equation}
\label{Fdiffusion}
\rho_{t+\tau}[\varphi;\psi]=\int \rho_{t}[\varphi+\eta;\psi]\gamma[\eta] D\eta,
\end{equation}
where $\gamma[\eta]$ is the probability functional of the variation $\eta$ in the state $\varphi$ and integration goes over all possible variations.

Before dealing with the integral over functions in (\ref{Fdiffusion}), 
let us express the conservation of particles under Brownian motion in differential form by means of the continuity equation. If $\rho_{t}({\bf a};{\bf b})$ is the density of particles at a point ${\bf a} \in \R^3$ and ${\bf j}_{t}({\bf a}; {\bf b})$ is the particle current density, then
\begin{equation}
\label{conti}
\frac{\partial \rho_{t}({\bf a};{\bf b})}{\partial t}+\nabla {\bf j}_{t}({\bf a};{\bf b})=0.
\end{equation}
Under the CQ-extension of the Brownian motion, (\ref{conti}) is replaced with the continuity equation that follows from the Schr{\"o}dinger dynamics. This is the same equation with 
\begin{equation}
\label{contiS}
\rho_{t}=|\psi |^2, \  \ {\rm and } \  \ {\bf j}_{t}=\frac{i\hbar}{2m}(\psi \nabla {\overline \psi}-{\overline \psi}\nabla \psi).
\end{equation}
In particular, for the states $\psi \in M^{\sigma}_{3,3}$ we obtain
\begin{equation}
\label{psiB}
{\bf j}_{t}=\frac{\bf p}{m} |\psi|^2={\bf v} \rho_{t}.
\end{equation}
Because the restriction of Schr{\"o}dinger evolution to $M^{\sigma}_{3,3}$ is the corresponding Newtonian evolution,  the function $\rho_{t}$ in (\ref{psiB}) must be the density of particles. This density was denoted earlier by $\rho_{t}({\bf a}; {\bf b})$. It gives the number of particles that start at ${\bf b}$ and by the time $t$ reach a neighborhood of the point ${\bf a}$.
The relation $\rho_{t}({\bf a}; {\bf b})=\rho_{t}[{\tilde \delta}^{3}_{\bf a}; {\tilde \delta}^{3}_{\bf b}]$ tells us that $\rho_{t}$ in (\ref{contiS}) must be then the state density
$\rho_{t}[{\tilde \delta}^{3}_{\bf a}; \psi]$. It gives the number of particles in initial state $\psi$ found under the measurement at time $t$ on a neighborhood of the point ${\bf a}$ in $\R^3$. 

The CQ-extension and the continuity equation already gave us the ``correct" form of the density of states functional: $\rho_{t}[{\tilde \delta}^{3}_{\bf a}; \psi]=|\psi({\bf a})|^2$. However, the dynamics of the system is missing: we know the density of states and therefore the relative frequency of transition of $\psi$ 
to ${\tilde \delta}^{3}_{\bf a}$ for different values of ${\bf a}$ (the Born rule), but how did the state get there?
In the classical model under discussion, the measured macroscopic particle undergoes a Brownian motion. Then, when the state of the particle is constrained to $M^{\sigma}_{3}=\R^3$, the CQ-extension must imply the usual diffusion on $\R^3$.
The restriction of (\ref{Fdiffusion}) to $M^{\sigma}_{3}$ means that $\varphi=\tilde{\delta}^{3}_{\bf a}$ and $\eta=\tilde{\delta}^{3}_{{\bf a}+{\bf \epsilon}}-\tilde{\delta}^{3}_{\bf a}$, where ${\bf \epsilon}$ is a displacement vector in $\R^3$. As we already know, the function $\rho_{t}[\tilde{\delta}^{3}_{\bf a};\tilde{\delta}^{3}_{\bf b}]=\rho_{t}({\bf a};{\bf b})$ is the usual density of particles in space.
Let us substitute this into (\ref{Fdiffusion}), replace $\gamma[\eta]$ with the corresponding probability density function 
$\gamma({\bf \epsilon})$ and integrate over the space $\R^3$ of all possible vectors ${\bf \epsilon}$. In the classical Brownian motion under investigation $\gamma({\bf \epsilon})$  is independent of ${\bf a}$ and the direction of ${\bf \epsilon}$ (space symmetry). Therefore, the terms $\int \epsilon^{k}\gamma({\bf \epsilon})d{\bf \epsilon}$ and  $\int \epsilon^{k}\epsilon^{l}\gamma({\bf \epsilon})d{\bf \epsilon}$ with $k \neq l$ vanish. It follows that
\begin{equation}
\label{diffusionR}
\frac{\partial \rho_{t}({\bf a};{\bf b})}{\partial t}=k\Delta \rho_{t}({\bf a};{\bf b}),
\end{equation}
where $k=\frac{1}{2\tau} \int \epsilon^2 \gamma({\bf \epsilon})d{\bf \epsilon}$ is a constant.

The derived diffusion equation (\ref{diffusionR}) describes the dynamics of an ensemble of particles in the classical space $M^{\sigma}_{3}$.
Because the particles start at a point, the initial distribution in (\ref{diffusionR}) is given by the delta function. By the time $\tau$ (typical time of measurement) the distribution is Gaussian and according to (\ref{contiS}) we must have  
\begin{equation}
\rho_{\tau}({\bf a};{\bf b})=|\tilde{\delta}^{3}_{{\bf b}}({\bf a}) |^2=\frac{1}{\sigma^3}|(\tilde{\delta}^{3}_{{\bf a}}, \tilde{\delta}^{3}_{{\bf b}})|^2,
\end{equation}
where the relationship $\tilde{\delta}^{3}_{{\bf a}}({\bf x})=\sigma^{\frac{3}{2}}\delta^3_{\bf a}({\bf x})$ was used. (See \cite{KryukovMacro}.) From (\ref{contiS}) we similarly get
\begin{equation}
\label{BornDerivedL1}
\rho_{\tau}[{\tilde \delta}^{3}_{\bf a}; \psi]=\frac{1}{\sigma^3}|(\tilde{\delta}^{3}_{{\bf a}}, \psi)|^2.
\end{equation}
A general initial state $\psi$ of a microscopic particle under observation can be always approximated as well as needed by a linear combination
\begin{equation}
\label{psiN}
\psi({\bf x})=\underset{\bf b}{\sum} C_{\bf b} \tilde{\delta}^{3}_{{\bf b}}({\bf x}).
\end{equation}
Substituting this into (\ref{BornDerivedL1}) and using the near-orthogonality of the states ${\widetilde \delta}^{3}_{\bf b}$ for different values of ${\bf b}$, we obtain
\begin{equation}
\label{psiL}
\rho_{t}[{\tilde \delta}^{3}_{\bf a}; \psi]=\underset{\bf b}{\sum} \left|C_{\bf b}\right|^{2} |\tilde{\delta}^{3}_{{\bf b}}({\bf a})|^{2}=\underset{\bf b}{\sum} \left|C_{\bf b}\right|^{2} \rho_{t}({\bf a};{\bf b}).
\end{equation}
Because $\rho_{t}({\bf a};{\bf b})$ satisfies (\ref{diffusionR}), which is a linear equation, we conclude that the state density $\rho_{t}[{\tilde \delta}^{3}_{\bf a}; \psi]$ also satisfies the equation. The Brownian motion experienced by the components $\tilde{\delta}^{3}_{{\bf b}}$ of state $\psi$ of the measured particle results in a ``Born-like" motion of $\psi$ itself. During the observation an ensemble of particles in the initial state $\psi$ diffuses in the space of states so that the resulting probability distribution to find the state at a point $\tilde{\delta}^{3}_{{\bf a}}$ in the classical space $M^{\sigma}_{3}$ is given by the Born rule.

Note that (\ref{BornDerivedL1}) can be written as a quadratic form in states $\tilde{\delta}^{3}_{{\bf b}}$. Namely, 
\begin{equation}
\label{BornDerivedL}
\rho_{\tau}[{\tilde \delta}^{3}_{\bf a}; \psi]=\int c({\bf x},{\bf y})\tilde{\delta}^{3}_{{\bf b}}({\bf x})\tilde{\delta}^{3}_{{\bf b}}({\bf y})d{\bf x}d{\bf y},
\end{equation}
where
\begin{equation}
\label{c0}
c({\bf x},{\bf y})=\frac{1}{\sigma^3}{\overline \psi}({\bf x})\psi({\bf y}).
\end{equation}
Extending (\ref{BornDerivedL}) from $M^{\sigma}_{3}$ to $L_{2}(\R^3)$ by linearity, we obtain
\begin{equation}
\label{BornDerivedL2}
\rho_{\tau}[\varphi; \psi]=\int c({\bf x},{\bf y})\varphi({\bf x}){\overline \varphi}({\bf y})d{\bf x}d{\bf y}.
\end{equation}
This gives the Born rule for probability of transition between arbitrary states
\begin{equation}
\label{c5}
\rho_{\tau}[\varphi; \psi]=\frac{1}{\sigma^3}|(\varphi, \psi)|^2.
\end{equation}
In particular, the probability of transition from state $\psi$ to state $\varphi$ is the same as the probability of transition from $\varphi$ to $\psi$. (This latter condition would also lead us from (\ref{BornDerivedL}) to (\ref{c5}).) We also see that the density of states is well-defined on the projective space $CP^{L_{2}}$ and depends only on the distance between the initial and the end states in the Fubini-Study metric on $CP^{L_{2}}$.

We started with a classical model of measurement and used properties of its CQ-extension to derive the diffusion of states for an ensemble of microscopic particles exposed to a position measurement. This leads us to the conclusion that  the probability of finding the ``measured" state $\psi$ at a point $\varphi$ depends only on the distance between the states in the Fubini-Study metric and is given by the Born rule. 

It is important that we could have started with an arbitrary classical model of measurement consistent with Newtonian dynamics in a ``noisy" potential $V({\bf x}, t)$ and with the normal distribution of position random variable. The existence and uniqueness of the CQ-extension would allow us to model interaction of a microscopic particle with the measuring device and, if space symmetry is preserved, to derive the Born rule dynamically. 
At the same time, whenever the outcomes of a classical measurement experiment satisfy the normal distribution law, the process of measurement can be interpreted in terms of diffusion. The distribution of outcomes must then satisfy the diffusion equation (\ref{diffusionR}), which makes the provided derivation of the Born rule valid.

The derivation means that under the CQ-extension of the classical Brownian motion the state of the particle is equally likely to be pushed in any direction in $CP^{L_2}$ (as the derived Born rule is direction-insensitive). The latter motion satisfies the standard, linear Schr{\"o}dinger equation for the particle in a fluctuating potential.

Equal likelihood  of all directions of displacement of state may seem surprising. Recall, however, that the Brownian particle is equally likely to be pushed in any spatial direction at any point in space $\R^3$. At the same time, the space $\R^3=M^{\sigma}_{3}$ is complete in $L_{2}(\R^3)$. Given a state $\psi$, we can decompose it with respect to the elements of $M^{\sigma}_{3}$. Because each component of $\psi$ is equally likely to be displaced in any direction in $M^{\sigma}_{3}$, the ratios of the coefficients of decomposition of state in the original basis can freely fluctuate. 

A random change in the modulus and phase of the coefficients is readily produced by various ``noises" in space that are typical when a quantum system interacts with the environment. Generic stochastic conditions imposed on the noise lead one to the same conclusion by means of examples. Note also that the equal probability of various directions of fluctuation of state is in agreement with the known result  stating that unitary evolution is unable to yield a concentration of the state in Hilbert space \cite{Percival}.

It seems at first that not much has been achieved. The fact that a noise may lead to random fluctuation of state is rather simple and goes against of what one normally tries to achieve when explaining collapse under measurement. The collapse models utilize various ad hoc additions to Schr{\"o}dinger equation with the goal of explaining why the state under a random walk ``concentrates" to an eigenstate of the measured observable (usually, position or energy) \cite{Pearle76}-\cite{Bassi}. Instead, we see a clear indication that under a generic measurement, the state has equal probability of moving in any direction and ``diffuses" isotropically into the space of states. Surprisingly, this diffusion on the space of states, {\it being in agreement with the unitary Schr{\"o}dinger evolution}, is capable of addressing the major issues of measurement in quantum theory.

In fact, under the diffusion, the probability of transition of $\psi$ to any other state $\varphi \in CP^{L_{2}}$ was shown to depend only on the distance between the states and to satisfy the Born rule. So diffusion that follows from the linear Sch{\"o}dinger evolution with fluctuating potential is capable of explaining transitions of quantum states. This major observation signifies that the role of the measuring device may be reduced to initiating the diffusion(creating a ``noise") and to registering a particular location of the diffused state. For instance, the ``noise" in the position measuring device under consideration is due to the stream of photons. The device then registers the state reaching a point in $M^{\sigma}_{3}$. In a similar way, a momentum measuring device registers the diffused states that reach the eigen-manifold of the momentum operator (which is the image of $M^{\sigma}_{3}$ under the Fourier transformation). 
It follows, in particular, that the measuring device in quantum mechanics is not responsible for creating a basis into which the state is to be expanded. If several measuring devices are present, they are not ``fighting" for the basis.  When the eigen-manifolds of the corresponding observables don't overlap, only one of them can ``click" for the measured particle as the state can reach only one of the eigen-manifolds at a time. 
Note also the similarity in the role of measuring devices in quantum and classical mechanics: in both cases the devices are designed to measure a particular physical quantity and inadvertently create a ``noise", which results in a distribution of values of the measured quantity. 

Coming back to the derivation of the diffusion equation  (\ref{diffusionR})  and the Born rule (\ref{BornDerivedL1}), we see how the Brownian motion experienced by the components $\tilde{\delta}^{3}_{{\bf b}}$ of state $\psi$ of the measured particle results in a ``Born-like" motion of $\psi$ itself. The probability density to find the state $\psi$  at a point $\tilde{\delta}^{3}_{{\bf a}}$ (particle at a point ${\bf a}$) is given by (\ref{psiL}). It is a weighted sum of the normal probability distributions for each components. The coefficients $C_{\bf b}$ identify the position of the initial point $\psi$ in the space of states $CP^{L_{2}}$ relative to the classical space $M^{\sigma}_{3}$. This initial position determines the conditional probability for the state of reaching a particular point $\tilde{\delta}^{3}_{{\bf b}}$ given that it has reached $M^{\sigma}_{3}$ as a result of diffusion.

Furthermore, due to linearity  of the diffusion equation (\ref{diffusionR}), the ``cloud" $|\psi({\bf a})|^{2}$ evolves in time as if it was in fact a cloud ``diffusing" in $\R^{3}$, rather than a single state (point) moving stochastically in $CP^{L_{2}}$. The reason for it is clear: since we restrict the outcomes of measurements to only those in the space $M^{\sigma}_{3}=\R^3$, the probability density is the probability of getting a particular position value in $\R^3$. The fact that the original state does not belong to $M^{\sigma}_{3}$ is not explicit in the density function $\rho_{t}({\bf a})$, giving us the confusing cloud interpretation in $\R^3$.

So what does it all say about measurement of position of macroscopic and microscopic particles? During the period of observation of position of a macroscopic particle in the model, the position random variable experiences a Brownian motion. Normally, observation happens during a short enough interval of time so that the particle does not get displaced much and the spread of the probability density is sufficiently small. A particular value of position variable during the observation is simply a realization of one of the possible outcomes. The Brownian motion of macroscopic particle can be equivalently thought of as either a stochastic process ${\bf b}_{t}$ with values in $\R^{3}$ or  a process $\tilde{\delta}^{3}_{{\bf b}, t}$ with values in $M^{\sigma}_{3}$. The advantage of the latter representation is that the position random variable $\tilde{\delta}^{3}_{\bf b}$ gives both, the position of the particle in  $M^{\sigma}_{3}=\R^{3}$ and the probability density to find it in a different location ${\bf a}$ (in the state $\tilde{\delta}^{3}_{{\bf a}}$), due to uncontrollable interactions with the surroundings under observation and the resulting Brownian motion.

Measuring position of a microscopic particle has, in essence, a very similar nature.  Under observation each component of the state $\psi$ of the particle in (\ref{psiN}) experiences the usual Brownian motion on $M^{\sigma}_{3}$ (or the phase space $M^{\sigma}_{3,3}$). As a result, the state $\psi$ itself becomes a random variable, taking values in the space of states $CP^{L_{2}}$. To measure position is to observe the state on the submanifold $M^{\sigma}_{3}$ (or $M^{\sigma}_{3,3}$) in $CP^{L_{2}}$. In this case, the random variable $\psi$ assumes one of the values $\tilde{\delta}^{3}_{{\bf a}}$,  with the uniquely defined probability density compatible with the normal density in the space $\R^3$. This probability density, associated with the conditional probability to find the state $\psi$ at $\tilde{\delta}^{3}_{{\bf a}}$ given that $\psi$ has reached $M^{\sigma}_{3}$, is exactly the one given by the Born rule. Here too the random variable $\psi$ gives both, the position of the state of the particle in $CP^{L_{2}}$ and the probability density to find the particle in a different state $\tilde{\delta}^{3}_{{\bf a}}$.

So the difference between the measurements is two-fold. First, under a measurement the state $\psi$ of a microscopic particle is a random variable over the entire space of states $CP^{L_{2}}$ and not just over the submanifold $M^{\sigma}_{3}$. Second, unless $\psi$ is already constrained to $M^{\sigma}_{3}$ (the case which would mimic the measurement of position of a macroscopic particle), to measure position is to observe the state that ``diffused" enough to reach the submanifold $M^{\sigma}_{3}$. To put it differently, the measuring device is not where the initial state was. Assuming the state has reached $M^{\sigma}_{3}$, the probability density of reaching a particular point in $M^{\sigma}_{3}$ is given, as we saw, by the Born rule.

We don't use the term collapse of position random variable when measuring position of a macroscopic particle. Likewise, there is no physics in the term collapse of the state of a microscopic particle. Instead, due to the diffusion of state, there is a probability density to find the particle in various locations on $CP^{L_{2}}$. In particular, the state may reach the space manifold $M^{\sigma}_{3}=\R^3$. If that happens and we have detectors spread over the space, then one of them clicks. If the detector at a point ${\bf a} \in \R^3$ clicks, that means the state is at the point $\tilde{\delta}^{3}_{{\bf a}} \in CP^{L_{2}}$ (that is, the state {\it is} $\tilde{\delta}^{3}_{{\bf a}}$). The number of clicks at different points ${\bf a}$  when experiment is repeated is given by the Born rule. The state is not a ``cloud" in $\R^{3}$ that shrinks to a point under observation. Rather, the state is a point in $CP^{L_{2}}$ which may or may not be on $\R^3=M^{\sigma}_{3}$. When the detector clicks we know that the state is on $M^{\sigma}_{3}$.

Note once again that there is no need in any new mechanism of ``collapse". There is no ``concentration" of state involved and the stochastic process is in agreement with the conventional Schr{\"o}dinger equation with a randomly fluctuating potential (``noise"). The origin of the potential depends on the type of measuring device or properties of the environment, capable of ``measuring" the system. Fluctuation of the potential
can be traced back to thermal motion of molecules, atomic vibrations in solids, vibrational and rotational molecular motion, and the surrounding fields. Transition from individual effect of a ``kick" on a spatial component of the state $\psi$ in (\ref{psiN}) to their combined effect on $\psi$ and the resulting stochastic process require a change in description: {\it The linear equation for the state results for the ensemble of states in a linear equation for the probability density, i.e., the diffusion equation, which, of course, is not linear in $\psi$}.

\section{Generalities of the classical behavior of macroscopic bodies}

It was demonstrated that the Schr{\"o}dinger evolution of state constrained to the classical phase space $M^{\sigma}_{3,3}$ results in the Newtonian motion of the particle. A similar result holds  true for systems of particles.  To reconcile the laws of quantum and classical physics, one must also explain the nature of this constraint. Why are microscopic particles free to leave the classical space, while macroscopic particles are bound to it? What is the role of decoherence in this, if any?

Suppose first that the macroscopic particle under consideration is a crystalline solid. The position of one cell in the solid defines the position of the entire solid. If one of the cells was observed at a certain point at rest, the state of the solid immediately after the observation (in one dimension) is the product
\begin{equation}
\label{tensorP}
\varphi=\tilde{\delta}_{a}\otimes \tilde{\delta}_{a+\Delta}\otimes \ ... \ \otimes \tilde{\delta}_{a+n\Delta},
\end{equation}
where $\Delta$ is the lattice length parameter.
The general quantum-mechanical state of the solid is then a superposition of states (\ref{tensorP}) for different values of $a$ in space:
\begin{equation}
\label{tensorSP}
\varphi=\sum_{a}C_{a}\tilde{\delta}_{a}\otimes \tilde{\delta}_{a+\Delta}\otimes \ ... \ \otimes \tilde{\delta}_{a+n\Delta}.
\end{equation}
Why would non-trivial superpositions of this sort be absent in nature? 

The classical phase space $M^{\sigma}_{3n,3n}$ of a $n$-particle system consists of all tensor products $\varphi_{1}\otimes  . . . \otimes \varphi_{n}$, defined up to a phase factor, with the state $\varphi_{k}$ of each particle given by  (\ref{del1}). As demonstrated in \cite{KryukovMacro} and discussed in this paper, the Schr{\"o}dinger dynamics of $n$-particle system constrained to $M^{\sigma}_{3n,3n}$ is the Newtonian dynamics of the system. In the section titled ``Extension of Newtonian dynamics to the space $CP^{L_{2}}$ of quantum states" applied to the case of $n$ particles it was also shown that there exists a unique unitary evolution whose restriction to $M^{\sigma}_{3n,3n}$ reproduces the motion of $n$ particles in Newtonian dynamics. This evolution is governed by the usual Hamiltonian $\sum_{k}\frac{{\widehat p}^{2}_{k}}{2m_{k}}+V({\bf x}_{1},...,{\bf x}_{n})$. 

Note also that the isomorphism $\omega_{n}: \R^3 \times...\times \R^3 \longrightarrow M^{\sigma}_{3n}$, $\omega_{n}({\bf a}_{1}, ... , {\bf a}_{n})=\tilde{\delta}^{3}_{{\bf a}_{1}}\otimes ... \otimes  \tilde{\delta}^{3}_{{\bf a}_{n}}$ allows us to interpret $n$-particle states in $M^{\sigma}_{3n}$ as positions of $n$ particles in the classical space $\R^{3}$. A similar map identifies the submanifold $M^{\sigma}_{3n,3n}$ with the classical phase space of $n$ particles. 
Both maps are particular instances of the Segre embedding. These maps together with the discovered relationships of the classical and quantum dynamics allow us to think of $M^{\sigma}_{3n}$ and $M^{\sigma}_{3n,3n}$ as the classical space and phase space with $n$ particles.

To understand the dynamics of macroscopic bodies under measurement, consider the Brownian motion of a crystalline solid. The motion of any solid can be represented by the motion of its center of mass under the total force acting on the body and a rotation about the center of mass. From the derivation of the Brownian motion applied to the solid one can see that the center of mass of the solid will experience the usual Brownian motion under the random force term, which is the  sum of forces acting from the surrounding particles on each cell. The related probability density must then satisfy the diffusion equation. 

These observations combined with the derivation in the section titled ``Observation of position of macroscopic and microscopic particles" allows us to derive the Born rule for the crystalline solid and explore its consequences. Suppose for simplicity that the solid is one-dimensional and consists of $n$-cells. (In this case the previous derivation for a single particle would work too, but the goal is to review what is involved when the measured system consists of many particles.) Let $\rho[\varphi;\psi]$  be the density of states functional on the space $CP^{L_{2}}$, where $L_{2}=L_{2}(\R)\otimes ... \otimes L_{2}(\R)$ is the tensor product of $n$ single particle Hilbert spaces. The conservation of states for the system reads as before
\begin{equation}
\label{FdiffusionN}
\rho_{t+\tau}[\varphi]=\int \rho_{t}[\varphi+\eta;\psi]\gamma[\eta] D\eta,
\end{equation}
where the meaning of terms is as in (\ref{Fdiffusion}). Define ${\tilde \delta}^{\otimes}_{a}=\tilde{\delta}_{a+\Delta_{1}}\otimes \tilde{\delta}_{a+\Delta_{2}}\otimes ... \otimes \tilde{\delta}_{a+\Delta_{n}} \in M^{\sigma}_{3n}$ and consider the functions
\begin{equation}
\rho_{t}(a;b)=\rho_{t}[ {\tilde \delta}^{\otimes}_{a}; {\tilde \delta}^{\otimes}_{b}],
\end{equation}
and 
\begin{equation}
\rho_{t}(a;\psi)=\rho_{t} [{\tilde \delta}^{\otimes}_{a}; \psi],
\end{equation}
where $a$, $b$ denote the center of mass and $\Delta_{k}$ describe the positions of each cell relative to the center of mass. From ({\ref{FdiffusionN}) we obtain the diffusion equation for $\rho_{t}(a;b)$ and the resulting Gaussian distribution for the position of the center of mass. Decomposition
$\psi=\sum_{b} C_{b} {\tilde \delta}^{\otimes}_{b}$
yields
\begin{equation}
\rho_{t}(a;\psi)=\sum_{b}|C_{b}|^2 \rho_{t}(a;b),
\end{equation}
which shows that $\rho_{t}(a;\psi)$ also solves the diffusion equation. The Born rule follows as before.

It is a well established and experimentally confirmed fact that macroscopic bodies experience an unavoidable interaction with the surroundings. 
Their ``cells" are pushed in all possible directions by the surrounding particles. For instance, a typical Brownian particle of radius between $10^{-9}m$ and $10^{-7}m$ experiences about $10^{12}$ random collisions per second with surrounding atoms in a liquid. The number of collisions of a solid of radius $10^{-3}m$ in the same environment is then about $10^{19}$ per second. Collisions with photons and other surrounding particles must be also added. Even empty space has on average about $450$ photons per $cm^3$ of space. 

The derivation of the Born rule for microscopic particles demonstrates that under a measurement (intentional or not) the state of a microscopic particle undergoes a diffusion on the space of states. On the other hand, as just discussed, macroscopic particles are always exposed to interaction with the surroundings. In a sense, their position is always measured. From this and the provided derivation of the Born rule, it follows that the state of a macroscopic particle always experiences a diffusion on the space of states rather than the free Schr{\"o}dinger evolution. (Note that the derivation of the Born rule given for a crystalline solid in one dimension can be clearly generalized to arbitrary solids in $\R^3$.)
As before, the underlying dynamics is the one given by the Schr{\"o}dinger equation. However, the collisions manifest themselves in a fluctuating potential term in the equation, which is present at all times and is responsible for the diffusion of state.

Now, suppose the state of a macroscopic body (in one dimension) is initially given by $\psi=\tilde{\delta}_{b+\Delta_{1}}\otimes ... \otimes \tilde{\delta}_{b+\Delta_{n}}$. Recall that this means that the initial distribution of position random variable is Gaussian with the center of mass at $b$. Under interaction with the surroundings the state $\psi$ undergoes a diffusion on the space of states. Consider the spatial (i.e., restricted to $M^{\sigma}_{3n}$) component of the diffusion.
As we know, the mean position of the center of mass will remain equal to $b$. 
Also, macroscopic bodies are distinguished by a sufficiently large number of ``cells" so that the total sum of forces from the collisions of cells with the surrounding particles at any time is almost exactly zero. 
In addition, the mass of the body increases with the number of cells, so the displacements generated by the total force become even smaller and for sufficiently large massive bodies can be safely disregarded. This means that the diffusion coefficient $k$ in (\ref{diffusionR}) vanishes so that the diffusion in space is trivial. But we know that the probability density of states is direction-independent: if the state does not diffuse in the space $M^{\sigma}_{3n}$, then it does not diffuse at all! Accordingly, the probability distribution remains constant in time. In the absence of additional potentials acting on the macroscopic body it will maintain its original state $\psi$. 

This qualitative analysis must be complemented by an accurate calculation. An estimate of fluctuations of state about the mean state $\psi$ in the classical space submanifold $M^{\sigma}_{3n}$ requires a selection of the parameter $\sigma$ and will be provided elsewhere. This estimate should also give us an idea about the boundary between classical and quantum world.

The situation is surprisingly similar to that of pollen grains and a ship initially at rest in still water. While under the kicks from the molecules of water the pollen grains experience a Brownian motion, the ship in still water will not move at all. Because of the established relation of Newtonian and Schr{\"o}dinger dynamics, this is more than an analogy. In fact, when the state is constrained to the classical phase space submanifold, the ``pushes" experienced by the state become the classical kicks in the space that lead to Brownian motion of the system.  

If now an external potential $V$ is applied to the macroscopic system, then according to (\ref{decomposition}), this will ``push" the state that belongs to the classical phase space submanifold in the direction tangent to the submanifold. Therefore, the external potential applied to a macroscopic body will not affect the motion of state in the directions orthogonal to the classical phase space submanifold. That means that the state will remain constrained to the submanifold. On the other hand, we know that the constrained state will evolve in accord with Newtonian dynamics in the total potential $V+V_{S}$, where $V_{S}$ is the potential created by the surroundings. However, since at any time $t$ the total force $-\nabla V_{S}$ exerted on the macroscopic body by the particles of the surroundings is vanishingly small, the body will evolve according to Newtonian equations with the force term $-\nabla V$. 

To be sure, the particles of the surroundings are responsible for friction. In the Hamiltonian description of interaction of the body with the surroundings the friction comes from a contribution to the total potential in the Hamiltonian. (See in particular the Hamiltonian of the Ullersma model in \cite{Sanz}.) However, whenever the friction can be neglected, the dynamics of the solid is determined by the force $-\nabla V$.

So far, the state of the system by itself was used as a dynamical variable, available during measurement. This was shown to be possible because dynamics of the measurement was a CQ-extension of the dynamics of a classical measurement, where the position variable is available. The constructed CQ-extension is consistent with the Schr{\"o}dinger dynamics. However, it is well known that interaction of the system with the measuring device and the environment creates an entangled state of the system and the surroundings. It is therefore impossible to talk about the state of the system by itself, contradicting the previous assertion. 

To understand the situation, let's begin once again with a measurement of the position of a macroscopic particle (i.e., the particle whose state is constrained to the classical phase space $M^{\sigma}_{3,3}$) by observing scattered photons or other particles. 
When the measured particle is observed this way, the incident particles and the measured particle exchange energy and momentum. The information about the measured particle "leaks out" into the environment, affecting potentially the entire universe. However, entanglement between the particle and the surroundings in the usual sense is absent. The state of the observed particle and the surroundings has the form
\begin{equation}
\label{deltaE}
\tilde{\delta}^3_{\bf a }\otimes E_{\bf a}, 
\end{equation}
where $E_{\bf a}$ represents the state of the apparatus and the environment. 
The state of the system belongs to the submanifold $M^{\sigma}_{3,3,E}$ of the tensor product of Hilbert spaces of the particle and the surroundings that consists of the product states (\ref{deltaE}).  The position of the particle is defined and can be found at any time, at least in principle. 
In some cases, the surroundings can be modeled by a potential and the position of the particle is found by solving Newton's equations of motion. In some cases to predict position of the particle, we have to consider a system consisting of the observed particle and particles in the surroundings. When many particles of the surroundings are involved, the position is best described stochastically. Typically, a diffusion equation can be used to find the probability distribution of the position random variable.

Suppose now the position of a microscopic particle is measured in the same way. Here too, in some cases interaction with the surroundings can be modeled by a potential. In some cases, we have to deal with a many-particle system and attempt solving the Schr{\"o}dinger equation for the system. For a large number of particles in the surroundings, the Schr{\"o}dinger equation results in a stochastic process. In the considered case this was shown to generate the diffusion of state on the space of states $CP^{L_{2}}$ of the particle. However, in the most general case, all we can claim is that the state $\Psi$ of the system consisting of the particle and the surroundings is a sum of terms in (\ref{deltaE}). In this case, the state of the system at any time is a point in the space of states $CP^{L_{2}}$, where $L_{2}$ is the tensor product of Hilbert spaces of the particle and the surroundings. 
As before, the random nature of interaction between the involved particles results in random fluctuations of $\Psi$ in the projective space $CP^{L_{2}}$ with equal probability of all directions of displacement in the tangent space $T_{\Psi}(CP^{L_{2}})$. The entangled state $\Psi$ undergoes a diffusion on $CP^{L_{2}}$. 
In particular, $\Psi$ can reach the submanifold $M^{\sigma}_{3,3,E}$ consisting of the product states $\tilde{\delta}^{3}_{\bf a} \otimes E_{\bf a}$. If that happens, the position of the state in $M^{\sigma}_{3,3}$ becomes defined. That is, the position of the particle in the classical sense is defined and can be recorded by the measuring device. As we saw in a similar derivation in this section, the conditional probability of $\Psi$ reaching a particular point in manifold $M^{\sigma}_{3,3,E}$, given that it has reached the manifold, satisfies the Born rule.

So far, decoherence \cite{Decoh} was not present in the discussion. 
Formally, decoherence is a mathematical expression of the fact that a quantum system interacting with the environment behaves like a probabilistic mixture and needs to be described by the probability and not by the state. 
In its ``pure" form it does not involve dynamical change in the components of the state.
The theory is centered around the issue of entanglement and the resulting loss of coherence. 
It does not usually go beyond recognition of the loss of coherence and the resulting need of probabilistic description of the system. It does not describe the way in which specific measurement results are obtained and does not derive the Born rule. 
At the same time, decoherence theory includes an array of very useful models that provide physical content for the theory. These models testify to the universal character of the loss of coherence and transition to classical probability resulting from interaction with the environment. 
Moreover, interaction with the incident particles in the model of spatial decoherence by scattering is what also triggers the diffusion of state under discussion here. In this context, decoherence may be considered a superficial expression of the underlying physical process of diffusion of state.
At any rate, the models of decoherence provide an additional support for a stochastic description of measurement and interaction with the surroundings. This is despite the characterization of diffusion models such as the one used here as ``fake" decoherence in the decoherence theory, due to their microscopically unitary character.


\section{Summary}

The dynamics of a classical $n$-particle mechanical system can be identified with the Schr{\"o}dinger dynamics constrained to the classical phase space submanifold $M^{\sigma}_{3n,3n}$ in the space of states. Conversely, there is a unique unitary time evolution on the space of states of a quantum system that yields Newtonian dynamics when constrained to the classical phase space.
This results in a tight, previously unnoticed relationship between classical and quantum physics. In particular, under a measurement of the position of a macroscopic particle, the position random variable obeys generically the normal distribution law. This implies under the extension the Born rule for transitions between quantum states. 
Therefore, any classical (i.e., based on Newtonian dynamics) model of measurement of a macroscopic particle that predicts the normal distribution of the position random variable extends in a unique way to the corresponding quantum (i.e., satisfying Schr{\"o}dinger dynamics) model that enforces the Born rule for probability of transition between states.  The central limit theorem makes it easy for the outcomes of a measurement of a classical system to satisfy the normal distribution law. It follows that the Born rule in measurements of a quantum system is as generic as the normal distribution law in classical measurements.

In the paper the proved relationship between classical and quantum concepts is taken to mean that physical laws that govern behavior of macroscopic and microscopic bodies are fundamentally the same. For instance, there exists a unique extension of the classical Brownian motion from the classical space submanifold $M^{\sigma}_{3}$ to the space of states $CP^{L_{2}}$ of the particle. Because the Brownian motion can model the process of measurement in classical physics, its unique extension is taken to be adequate for the description of measurements on microscopic systems.

With this understood, the process of measurement on a quantum system can be described in terms of a diffusion of state of the measured system in which the state has equal probability of being displaced in any direction in the space of states $CP^{L_{2}}$. The role of the measuring device reduces then to creating a ``noise" that triggers the diffusion in $CP^{L_{2}}$ and in recording the diffused state when it reaches a particular region in $CP^{L_{2}}$. The conclusion is that the so called collapse of the wave function in the framework is not about an instantaneous  ``concentration" of state near an eigenstate of the measured observable. Instead, it is about diffusion on the space of states under interaction with the measuring device and the environment. The ``collapse" to an eigenstate of an observable happens when the state under the diffusion reaches the eigen-manifold of that observable. In case of position measurements the state has to reach the classical space or phase space submanifolds in $CP^{L_{2}}$. Due to the enormous amount of collisions between a macroscopic body and the particles in the surroundings, position of the body is constantly measured. As a result, the diffusion process for macroscopic bodies can trivialize, which may explain why they stay in the classical space and therefore have a definite position.

We see that macroscopic and microscopic particles may not be so different after all. The only important distinction is that microscopic systems within the proposed framework live in the space of states while their macroscopic counterparts live in the classical space submanifold. Since our own life happens primarily in the macro-world, it is hard for us to understand the infinite-dimensional quantum world around us. As soon as the classical-space-centered point of view is extended to its Hilbert-space-centered counterpart, the new, clearer view of the classical-quantum relationship emerges.

\section{Acknowledgements}

I would like to express my sincere gratitude to Larry Horwitz for his unfailing interest, encouragement and support and for lively discussions of topics of common interest.


\end{document}